\documentstyle[preprint,aps]{revtex}
\begin{document}
\draft
\preprint{UM-P-93/42}
\title{CAN THE ELECTRIC CHARGES OF ELEMENTARY PARTICLES
CHANGE WITH
TIME?}

\author{A.Yu.Ignatiev and G.C.Joshi}
\address{Research Centre for High Energy Physics, School of
Physics, University of Melbourne,
Parkville 3052, Victoria, Australia}
\date{May 4, 1993}
\maketitle
\begin{abstract}
We present the limits on possible time variation of the
electric
charges of quarks and leptons at the time of primordial
nucleosynthesis within a model with dequantized electric
charges which allows for the case of charged neutrino and
neutron.

12.90+b, 95.30Cq
\\
\begin{center}
(Physical Review D, in press)
\end{center}
\\
\\
\\
\\
\begin{center}
email: sasha@tauon.ph.unimelb.edu.au (A.Yu.Ignatiev)
\\
email: joshi@bradman.ph.unimelb.edu.au (G.C.Joshi)
\end{center}
\end{abstract}
The assumption of spacetime invariance of fundamental
constants is one of the cornerstones of modern physics. Any
evidence for their space or time dependence would imply
dramatic consequences for cosmology, astrophysics,
geophysics
and physics as a whole (for a review see \cite{reviews}).

{}From the theoretical standpoint, there are several reasons
to expect that (at least some of) the fundamental constants
may in fact vary.

Chronologically the first one was Dirac's observation of the
fact that the ratio of Coulomb and gravitational interaction
of two protons is very close to the age of the Universe
measured in "nuclear" units of time. Dirac suggested that
the reason why the gravity is so weak is because the
Universe is very old \cite{Dirac}. In other words, according
to this "Large Number Hypothesis" the gravitational constant
decreases inversely proportional to time which means that in
the Early Universe gravity was as strong as
electromagnetism.

Quite different motivation for possible variability of
constants appeared more recently within Kaluza-Klein
theories \cite{Kaluza-Klein}. Here, the values of coupling
constants are determined by the radius of compactified
dimensions which can naturally depend on time thus causing
the time evolution of coupling constants.

In this paper we suggest to ask a related but quite a
different question: what would be the consequences of and
the limits on possible variation of {\em electric
charges\/} of elementary particles (while the values of
{\em coupling constants\/} are assumed to be strictly
conserved).
This is obviously more interesting (and more complicated)
pattern than the case of varying $\alpha$ because we allow
to
vary not only the electric charges but also their ratios. In
particular, we face an intriguing possibility that neutrino
and neutron might be charged in some distant past or, even
bolder, that the electric charges of all quarks and leptons
(including neutrinos) were equal to each other in the early
stages of the universal evolution.

To start discussing this problem we first need to choose
some model.

First of all we notice that the minimal standard model (MSM)
is
not flexible enough for this purpose  because even if we
vary any set of its coupling constants (or even all of them)
in an arbitrary manner, the ratios of the electric charges
of any two particles would still be time independent (e.g.
the neutrino and neutron charges would always be zero).

This is of course due to the fact that particle charges in
MSM obey the formula $Q=T_{3}+Y/2$ which does not contain
any free parameter capable of time variation. The same is
true also for any other model in which the electric charges
of quarks and leptons are quantized.

Another reason why the MSM is unsuitable for our purposes is
the masslessness of the neutrinos in MSM. If we wish to
consider  neutrinos with non-zero values of the electric
charge in the distant past we run into well-known
difficulties of the theories with massless charged particles
, a subject  which we are not going to address here.

Therefore the model we need must satisfy two requirements:
first, the electric charge must be {\em dequantized\/} (so
that the charge assignment contains a new parameter capable
of time variation) and second, the neutrinos must be massive
(the neutrino mass should be of course Dirac since the
Majorana mass is incompatible with the hypothesis of
neutrino charge).

The simplest possibility is to add a right-handed neutrino
to MSM and then ensure that it gets an affordable Dirac
mass. The recently emerged approach \cite{Melb} to the
problem of
electric charge quantization (see also \cite{I} )shows that
in
such model the electric charge is indeed dequantized through
the following form:

\begin{equation}
Q=T_{3} + Y/2+ \epsilon (B-L)
\end{equation}
where Y is the standard hypercharge, B and L are baryon and
lepton numbers defined in the standard way (B=1/3 for
quarks), epsilon being our free parameter. It can be shown
\cite{Melb} that in the model under consideration, B-L is
anomaly free symmetry so that the above definition of charge
does
not lead to any theoretical inconsistencies. As long as
$\epsilon$ is kept very small we do not contradict the
experiment either.

As was shown in \cite{Melb} the strongest limit on
the present value of $\epsilon$ comes from the data on the
neutrality of the neutron \cite{neutron}:$\epsilon < 10^{-
21}$.
This is certainly a very small number which creates a new
hierarchy problem of particle physics in addition to the
existing ones. Thinking about it
in the spirit of Dirac's hypothesis it is natural to
contemplate the situation when $\epsilon$ was much larger at
the earlier stages of the Universe evolution. Note that
$\epsilon =-1/2$ corresponds to the fascinating case of
"electric charge democracy" (i.e. the electric charges of
quarks and leptons, including neutrinos, and also of the
proton and neutron are all equal to $\pm 1/2$).

Primordial nucleosynthesis provides us with the unique
opportunity to measure the values of various fundamental
constants 1-100 sec after the Big Bang \cite{nuclsynt}.

Very briefly, primordial nucleosynthesis proceeds in three
stages: at the first stage the ratio of neutron to proton
concentration follows the equilibrium law due to weak p-n
transitions,  at the second stage (starting after about
$T=0.7 MeV$) this ratio gets frozen so that protons and
neutrons behave as almost free particles, with neutrons
decaying, and at the final stage the remaining neutrons get
bound with protons to form the light nuclei such as $^3 He$,
$^4He$, D, and $^7Li$ (for a recent review see,
e.g.\cite{nuclsyntrev}).

Since nonzero value of epsilon means that neutrino and
neutron get charged, and the charges of proton and electron
get changed, it can affect all of the three stages of the
nucleosynthesis in a variety of ways.

Decrease of neutrino decoupling temperature down to or
lower than the electron-positron annihilation temperature
(about 0.5 MeV) which would mean that neutrinos get heated
as a result of $e^{+}e^{-}$ annihilation, that is the
concentration
of neutrinos participating in weak transitions between n and
p gets larger which lowers the value of n/p freezing
temperature
$T_{f}$ and therefore n/p ratio and the abundancy Y of
primordial $^4He$.

Radiative corrections and possible appearance of Coulomb
barriers in weak interactions (due to nonzero charges
acquired by
neutron and neutrino) change the weak cross-sections which
again can
influence $T_{f}$, n/p ratio and Y($^4He$).

During the period of free neutron decay new radiative
corrections to
the neutron lifetime could affect the number of neutrons
left over and
therefore $Y(^4He)$.

At the stage of nuclear interactions the change of Coulomb
barriers
would affect nuclear cross-sections and therefore the yield
of all
isotopes, the heaviest ones being influenced most of all.
Besides, the
binding energies of nuclei would change. In the case of
positive
$\epsilon$ , this could result in possible total
destabilization
of some
nuclei which could lead to drastic changes in the whole
reaction
chain. On the other hand, in the case of negative $\epsilon$
,
possible
stabilization of some nuclei could also undermine the normal
process
of nucleosynthesis.

Let us consider all these factors in turn. (Recall that we
assumed all
coupling constants ($\alpha, G_{F}, \alpha_{S}$ etc.) do not
change
with time). We start by
considering the effect of positive $\epsilon$ on Coulomb
barriers in nuclear reactions.

To get the strongest bound on $\epsilon$ we need to consider
the reactions for the synthesis of the (stable) nucleus with
the highest charge and mass, that is $^7Li$.

$^7Li$ can be produced in two reactions:

at low baryon density ($n_{B}/n_{\gamma} < 3.10^{-10}$)
through the
fusion $^4He(^3H, \gamma)^7Li$ whereas at higher densities
through the
electron capture $^7Be(e^{-}, \nu_{e})^7Li$, $^7Be$ being
produced via
$^4He(^3He, \gamma)^7Be$.
The reaction rates are proportional to \cite{WFH}:
\begin{equation}
A^{-1/3}(Z_{1\epsilon}Z_{2\epsilon})^{1/3}T^{-
2/3}_{9}S_{eff}exp(-
\tau)
\end{equation}
where $A$ is the reduced mass number, $T_{9}$ is the
temperature in
$10^{9} K$,
\begin{eqnarray}
S_{eff} \approx
S(0)[1+0.09807(Z_{1\epsilon}Z_{2\epsilon})^{-2/3}A^{-
1/3}T_{9}^{1/3}] \\
\tau=4.249(Z_{1\epsilon}Z_{2\epsilon})^{2/3}A^{1/3}T_{9}^{-
1/3}\\
Z_{1\epsilon,2\epsilon}=Z_{1,2}+\epsilon A_{1,2}
\end{eqnarray}
For the reaction $^4He(^3H, \gamma)^7Li$ $S(0)=6.4.10^{-2}$
\cite{FCZ}.

In view of the well-known ambiguity in determining the
primordial abundancy of $^7Li$, we assume conservatively
that
the suppression factor for $^7Li$ production cannot be
larger
than 10. From this we deduce our upper bound on positive
$\epsilon$: $\epsilon < 0.07$.

As for the case of negative $\epsilon$ the neutron gets
negatively charged while the proton charge decreases which
leads to increasing the binding energies of the nuclei. We
assume that the total stabilization of $^7Be$ would
unacceptably distort the normal process of nucleosynthesis.

This results in the following constraint on negative
$\epsilon$:$\epsilon > -0.25$.

Using the results of Ref.\cite{corrections} it can be shown
that if these bounds hold, all the other
above listed effects can be neglected.

It is not unreasonable to expect that the evolution of
$\epsilon$ can be related in some way to the rate of the
universal expansion, that is to assume that the time
dependence of $\epsilon$ is given by, say, power law in the
Hubble constant H:$\epsilon(t) = const.H^{x}$.
Then, our bound on $\epsilon$ can be translated into the
bound
on the exponent x: $x < 1.5$.
It is instructive to compare this bound on x with the bounds
which can be obtained from the analyses of Oklo natural
nuclear reactor \cite{Oklo} and the ratio of$^{187}Re$ to
$^{187}Os$
abundancies \cite{Dyson}.
These data give very strong constraints on the variation of
$\alpha$ which can be translated into the following
constraints on
$\epsilon$ at the time $O(10^{9})y$:
Oklo: $\epsilon < 10^{-8}$; $^{187}Re$: $\epsilon < 10^{-5}$
However, these lead only to very loose constraints on the
exponent x: Oklo: $x < 130$; $^{187}Re$: $x < 160$.

This is due to the fact that although the nucleosynthesis
constraint is much weaker than these two, the epoch of
nucleosynthesis is only 100 sec after the Big Bang and the
Hubble constant then was much larger than it is at present.

Another constraint on the value of $\epsilon$ during
nucleosynthesis can be derived if one assumes  that the
proton-neutron mass difference varies proportionally to the
variation of $\epsilon$:$m_{n}-m_{p}=1.293 (1 + k\epsilon)
MeV$.

Since there seems to be no reliable way to calculate the
coefficient $k$, strictly speaking, we can give the
constraint only on
the
product $k\epsilon$ which follows from the requirement
$\Delta Y(^4He)
< 0.02$: $k\epsilon < 0.04$.

However, an estimate of the coefficient $k$ can be obtained
within
the quark model if one uses the following ansatz for the
electromagnetic contribution to the nucleon masses
\cite{Gasser,quarks}:
\begin{equation}
M^{\gamma}=\alpha \langle 1/r \rangle
(Q_{1}Q_{2}+Q_{1}Q_{3}+Q_{2}Q_{3})+C(Q_{1}^{2}+Q_{2}^{2}+Q_{
3}^{2})
\end{equation}
where $Q_{i}$ are the quark charges, $ \langle 1/r \rangle $
is the
average inverse distance of the two quarks in a baryon.
According to different estimates, $ \langle 1/r \rangle =
330/390 MeV
$ \cite{r1} or $ \langle 1/r \rangle = 255 MeV $ \cite{r2}.
As for the constant $C$, following \cite{Gasser} we will
assume
conservatively that
it can take any value in the range $0< C < 1.5 MeV$.
Now, we are able to obtain our estimate for $k$: $-2.2 < k <
-1$,
which means that, conservatively, $|\epsilon| < 0.04$.

The authors are grateful to X.-G.He and R.R.Volkas for
interesting discussions.

This work was supported in part by the Australian Research
Council.


\begin{references}
\bibitem{reviews}F.J.Dyson, in Current Trends in the Theory
of Fields, a
Symposium in Honor of P.A.M.Dirac,  ed. J.E.Lannutti and
P.K.Williams
(New York: AIP, 1978), p.163; F.J.Dyson, in Aspects of
Quantum Theory,
ed.A.Salam and E.P.Wigner, (Cambridge U. Press, 1972),
p.213;
E.R.Cohen, in Gravitational Measurements, Fundamental
Metrology and
Constants, ed. V.De Sabbata and V.N.Melnikov,
(Dordrecht:Kluwer
Academic Publ., 1988),p.91
\bibitem{Dirac} P.A.M.Dirac, Nature {\bf 139}, 323 (1937);
Proc.Roy.Soc.London, Ser.A {\bf 165}, 198 (1938)

\bibitem{Kaluza-Klein}Modern Kaluza-Klein Theories, ed. by
T.Appelquist,
A.Chodos and P.G.O.Freund (Addison-Wesley, Reading, MA,
1987)
\bibitem{Melb} R.R.Foot,G.C.Joshi, H.Lew and R.R.Volkas,
Mod.Phys.Lett. {\bf A5}, 95 (1990); {\em ibid.\/} {\bf A5},
2721
(1990);
X.-G.He, G.C.Joshi, H.Lew and R.R.Volkas,
Phys.Rev.{\bf D43}, R22 (1991); {\em ibid.\/} {\bf D44},
2118 (1991);
X.-G.He, G.C.Joshi and B.H.J.McKellar, Europhysics Lett.
{\bf 10},
709 (1989);
K.S.Babu and R.N.Mohapatra, Phys.Rev.Lett. {\bf 63}, 938
(1989);
Phys.Rev. {\bf D41}, 271 (1990); {\em ibid.\/} {\bf D42},
3866 (1990);
N.G.Deshpande, Oregon Report OITS-107 (1979) (unpublished);
R.R.Foot, H.Lew and R.R.Volkas, J.Phys.G {\bf 19}, 361
(1993);
E.Takasugi and M.Tanaka, Phys.Rev. {\bf D44}, 3706 (1991);
Progr.Theor.Phys. {\bf 87}, 679 (1992);

\bibitem{I}A.Yu.Ignatiev, V.A.Kuzmin and M.E.Shaposhnikov,
Phys.Lett.
{\bf B84}, 315 (1979); M.I.Dobroliubov and A.Yu.Ignatiev,
Phys.Rev.Lett. {\bf 65}, 679 (1990)
\bibitem{neutron}J.Baumann et al. Phys. Rev. {\bf D37},
3107(1988)

\bibitem{nuclsynt}
E.W.Kolb, M.J.Perry and T.P.Walker, Phys. Rev. {\bf D33},
869(1986);
P.Khare, Phys. Rev. {\bf D34}, 1936 (1986);
V.V.Dixit and M.Sher, Phys. Rev. {\bf D37}, 1097 (1988);
J.M.Irvine and R.J.Humphreys, Class.Quantum Grav. {\bf 3},
1013
(1986);
R.J.Scherrer and D.N.Spergel, Preprint OSU-TA-6/92
\bibitem{corrections}D.A.Dicus et al, Phys. Rev. {\bf 26},
2694
(1982)
\bibitem{nuclsyntrev} K.A.Olive et al CFA-89-2985 UMN-TH-
816/89;
A.M.Boesgaard and  G.Steigman, Ann.Rev.Astron.Astrophys.
{\bf 23},
319 (1985)
\bibitem{WFH}J.N.Bahcall, Ap.J. {\bf 143}, 259
(1966);R.Wagoner,
W.Fowler and F.Hoyle, Ap.J. {\bf 148}, 3 (1967)
\bibitem{FCZ}W.Fowler, G.Caughlan and B.Zimmerman,
Ann.Rev.Astron.Astroph. {\bf 5}, 525 (1967)
\bibitem{Oklo}A.I.Shlyakhter, Nature, {\bf 264}, 340 (1976);
ATOMPKI Report A/1
(1983)
\bibitem{Dyson}F.J.Dyson, Phys.Rev.Lett.{\bf 19}, 1291
(1967)
\bibitem{quarks}Y.Miyamoto, Progr.Theor.Phys. {\bf 35}, 175
(1966);
A.De Rujula, H.Georgi and S.Glashow, Phys.Rev. {\bf D12},
147
(1975);K.Lane and S.Weinberg, Phys.Rev.Lett. {\bf 37}, 717
(1976)
\bibitem{r1}C.Itoh et al., Progr.Theor.Phys. {\bf 61}, 548
(1979)
\bibitem{r2}N.Isgur, Phys.Rev. {\bf D21}, 779 (1980)
\bibitem{Gasser}J.Gasser and H. Leutwyler, Phys.Rep. {\bf
87}, 77,
(1982)
\end{references}
\end{document}